\documentclass{WileyMSP-template}

\usepackage{amssymb}

\begin{document}

\pagestyle{fancy}
\rhead{\includegraphics[width=2.5cm]{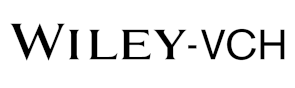}}

\title{In situ study of zinc peroxide decomposition to zinc oxide by X-ray absorption spectroscopy and reverse Monte-Carlo simulations}

\maketitle


\author{Alexei Kuzmin*},
\author{Inga Pudza},
\author{Konstantin Klementiev}

\dedication{}

\begin{affiliations}
	
Dr.\ A.\ Kuzmin,  I.\ Pudza \\
Institute of Solid State Physics, University of Latvia, Kengaraga street 8, LV-1063 Riga, Latvia\\
Email Address: a.kuzmin@cfi.lu.lv \\

Dr.\ K.\ Klementiev\\
MAX IV Laboratory, Lund University, PO BOX 118, SE-22100 Lund, Sweden

\end{affiliations}



\begin{abstract}
The Zn K-edge X-ray absorption spectroscopy has been used to investigate in situ the decomposition of zinc peroxide (ZnO$_2$) to zinc oxide (ZnO).  Principal component and linear combination analyses of the EXAFS spectra have been employed to identify the phase composition of the oxide upon heating to 900$^\circ$C. 
Only the ZnO$_2$\ phase has been found up to 180$^\circ$C, whereas only the nanocrystalline ZnO phase has occurred above  250$^\circ$C. Detailed structural information on the temperature dependence of the local environment of zinc atoms has been obtained using the reverse Monte Carlo simulations. A strong increase of disorder has been found upon approaching the decomposition temperature, evidenced by the broadening of Zn--O and Zn--Zn pair distribution functions and related mean-square relative displacements.
\end{abstract}


\section{Introduction}
\medskip
X-ray absorption spectroscopy is a powerful technique for in situ and operando investigations of  the local environment in materials with varying degrees of crystallinity and disorder \cite{Bordiga2013,Kuzmin2014,Mino2018,Grunert2020,Ghigna2021,Timoshenko2021review}.
Good sensitivity of the method to the short-range order around a selected element makes it useful to track the phase transitions in complex compounds \cite{Arai2013,Boccato2020,PUDZA2021102607,Miyanaga2021,Hao2021,Dua2021}. 
Technically, this means that, firstly, it is necessary to determine the type of phase(s) and then extract the structural information using a proper method of analysis \cite{Timoshenko2021review, Kuzmin2020}. 
Phase identification can be performed by comparison of the experimental data with a set of preselected reference compounds using principal component analysis (PCA) and/or linear combination analysis (LCA) procedure  \cite{Ressler2000,PANFILI2005}. The PCA method allows one to determine the minimum number and type of probable phases present in the sample, whereas their amount can be further quantified by the LCA. This approach will be used here to track the phase composition during the decomposition of zinc peroxide to zinc oxide.  

\medskip
A more refined analysis of X-ray absorption spectra can be performed based on ab initio multiple-scattering (MS) theory \cite{Filipponi1991,Rehr2000,Filipponi1995a,Natoli2003,DICICCO2020} to extract detailed structural information. 
If one is not limited to only the first coordination shell of the absorbing atom, then properly restricted MS expansion over a selected number of scattering paths is widely used \cite{Rehr2000,Zabinsky1995,Rehr2009}. 
While such an approach is commonly adopted, it has a drawback when the analysis is extended to the outer coordination shells due to the rapidly increasing number of parameters \cite{Kuzmin2014}.  As a result, a correlation between the parameters often  leads to their nonphysical values despite a good agreement between the calculated and experimental extended X-ray absorption fine structure (EXAFS) spectra. To overcome this problem, constraints can be used on the correlation between the parameters or the range of their valid values. A natural way to address this issue is to employ the reverse Monte Carlo (RMC) method choosing the starting model based on diffraction data, i.e., lattice parameters, and allowing only small atomic displacements during the simulation  \cite{Timoshenko2012rmc,Timoshenko2014}. 
Such an approach maintains the average structure of the material allowing to account for thermal and weak static disorder within ab initio MS formalism \cite{TIMOSHENKO2014194}.  Here, it will be demonstrated by analyzing the temperature-dependent Zn K-edge EXAFS spectra for zinc peroxide and zinc oxide as an example. 

\medskip
Zinc oxide exists at ambient conditions in two phases -- wurtzite-type ZnO \cite{Ozgur2005,Janotti2009} and peroxide ZnO$_2$\ \cite{chen2009synthesis,Bocharov2022}. In both structures, oxygen atoms compose the first coordination shell of zinc. However, the coordination of zinc atoms is tetrahedral in ZnO, while it is octahedral in ZnO$_2$. A peculiar crystallographic structure of ZnO$_2$\ containing peroxide (O$_2$)$^{2-}$\ groups is responsible for its low decomposition temperature of about 200-230$^\circ$C \cite{uekawa2003nonstoichiometric,bergs2016ultrasmall}, which can be further affected by reducing the size of ZnO$_2$\ crystallites \cite{chen2009synthesis,Daley2013}. 
Upon decomposition, ZnO$_2$\ transforms to nanocrystalline ZnO (nano-ZnO). It was suggested in \cite{Daley2013} that a formation of an intermediate amorphous phase can occur before the growth of the nano-ZnO phase.
The use of ZnO$_2$\ as a precursor for the production of ZnO nanoparticles and thin films has been demonstrated in several studies \cite{uekawa2003nonstoichiometric,Daley2013,zhang2005low,han2008properties,RAWAT2018,Daley2021}.
Among many applications \cite{Bocharov2022}, such nanoparticles may be of interest in printed optoelectronics as part of zinc oxide inkjet inks \cite{Sharma2017,Arrabito2020,Morales2021}. 

\medskip
The process of ZnO$_2$ decomposition to ZnO has recently been studied  in the temperature range from ambient to 500$^\circ$C by in situ combined X-ray diffraction (XRD) and X-ray absorption spectroscopy in \cite{Daley2021}. The results obtained were supplemented by thermogravimetric analysis (TGA). The weight loss observed in TGA and the change of the ZnO$_2$ (022) reflection in the XRD patterns indicated the decomposition of ZnO$_2$ and the formation of ZnO  in the temperature range from 175 to 250$^\circ$C \cite{Daley2021}. It was also proposed in \cite{Daley2021} from the analysis of XRD patterns that a poorly crystalline (amorphous like) ZnO phase exists in the temperature range of about 250-350$^\circ$C, while above 350$^\circ$C it becomes more ordered.  At the same time, an analysis of the Zn K-edge X-ray absorption near edge structure (XANES) and EXAFS suggested that a large number of defects are present in the ZnO obtained in this way \cite{Daley2021}.

\medskip
In this study, we performed a high-temperature (20-900$^\circ$C) Zn K-edge X-ray absorption spectroscopy study of two zinc oxide phases (ZnO and ZnO$_2$) to probe their temperature-dependent local structure and lattice dynamics and to elucidate the mechanism of ZnO$_2$-to-ZnO decomposition. The PCA method was used to determine the number of phases that exist at each temperature, while the LCA method was used to determine the type of each phase.  As a result, the temperature ranges of the existence and coexistence of the two phases were established. This information was used to analyze the local atomic structure of each phase in detail using RMC simulations.

\section{Data Analysis}

\medskip
Selected in situ experimental Zn K-edge  XANES and EXAFS spectra are shown in \textbf{Figure\ \ref{fig1}}. The EXAFS spectra were extracted following the conventional procedure \cite{Kuzmin2014}.   

\medskip
The analysis of the extracted Zn K-edge EXAFS spectra was performed using the PCA method to evaluate the phase composition of the sample, which changes upon the ZnO$_2$\ decomposition. 
Two main components corresponding to ZnO$_2$\ and wurtzite-type ZnO phases were identified.  
The amount of the ZnO$_2$\ fraction obtained by the LCA method as a function of temperature is shown in \textbf{Figure\ \ref{fig2}}. 
As one can see, two zinc oxide phases coexist in the temperature range of 190-240$^\circ$C, whereas only the ZnO$_2$\ phase is present in the sample up to 180$^\circ$C, and only the ZnO phase exists above 250$^\circ$C. Therefore, the EXAFS spectra acquired in these two temperature ranges were analyzed further in detail based on the crystallographic structure of the respective zinc oxide phase using the RMC simulations
\cite{Timoshenko2012rmc,Timoshenko2014}. The EvAX code \cite{evax2017} implementing 
the RMC method with the evolutionary algorithm approach was used. The method was previously applied to the analysis of the low-temperature EXAFS spectra of microcrystalline ZnO in \cite{TIMOSHENKO2014194,Timoshenko20141472} and demonstrated good sensitivity to small variations of the local atomic structure upon the lattice thermal expansion.

\medskip
To perform RMC simulations, an initial model of the oxide structure (a simulation box) was constructed  as a supercell with a size of 4$a$$\times$4$a$$\times$4$a$\ for ZnO$_2$\ or 
4$a$$\times$4$a$$\times$2$c$\ for ZnO based on the crystallographic parameters of the unit cell obtained by X-ray ($a$=4.896~\AA\ for ZnO$_2$\ \cite{Bocharov2022}) or neutron ($a$=3.250~\AA\ and $c$=5.204~\AA\ for  ZnO \cite{Albertsson1989}) diffraction. 
Periodic boundary conditions were employed to avoid problems with boundary effects. 
The size of the supercell was selected to be large enough to avoid artificial correlation effects and to collect reasonable statistics during acceptable calculation time. 
The number of atoms in the supercell was equal to the stoichiometric one for both oxide phases.
Several RMC simulations with different pseudo-random number sequences were performed to accumulate even better statistics. The Zn K-edge EXAFS spectra were calculated for each zinc atom in the supercell and averaged to obtain the configuration-averaged (CA) EXAFS spectrum.  
At each step of the RMC simulation, all atoms in the supercell were randomly displaced with the maximum allowed displacement of 0.4~\AA\  to account for thermal and/or static disorder.  The difference between the Morlet wavelet transforms (WTs) of the experimental and CA-EXAFS $\chi(k)k^2$\ spectra was used as a criterion for optimization \cite{Timoshenko2009wavelet}. 
The use of the WT guarantees the agreement between experiment and theory simultaneously in $k$\ and $R$\ spaces. 

\medskip
Ab initio real-space multiple-scattering FEFF8.50L code \cite{Rehr2000,Ankudinov1998} 
was used to calculate theoretical CA-EXAFS spectra. 
The EXAFS amplitude damping due to the photoelectron inelastic losses was accounted for using the complex exchange-correlation Hedin–Lundqvist potential \cite{Hedin1971}, and the EXAFS amplitude reduction factor $S_0^2$\ was set to 1.0 \cite{Kuzmin2014}.
Both single-scattering (SS) and multiple-scattering (MS) contributions were taken into account. 
MS contributions included all scatterings up to the 8th order. A large number of scattering paths was reduced by grouping similar ones and including  into the analysis only important paths with the relative amplitude of corresponding partial contribution to the total EXAFS greater than 0.1-1\% \cite{Timoshenko2014}. As a result of such path selection, only 60 and 43 scattering paths were involved in the cases of ZnO$_2$\ and ZnO, respectively.

\medskip
Comparison of the experimental and RMC calculated Zn K-edge EXAFS spectra $\chi(k)k^2$
for selected temperatures is shown in \textbf{Figures\ \ref{fig3}} and \textbf{\ref{fig4}} in the $k$\ and $R$\ spaces. 
Note that the peaks in the Fourier transforms (FTs) are located at distances that are slightly shorter than their crystallographic values because the FTs were not corrected for the phase shift present in the EXAFS equation \cite{Kuzmin2014}. 
The coordinates of atoms in the final RMC simulation box were used to calculate  pair distribution functions  (PDFs) $g_{\rm Zn-O}$\ and $g_{\rm Zn-Zn}$\ and  bond angle distribution functions (BADFs) for the O--Zn--O angles  (\textbf{Figure\ \ref{fig5}}).

\section{Results and Discussion}
\begin{figure*}[h]
	\centering
	\includegraphics[width=0.7\linewidth]{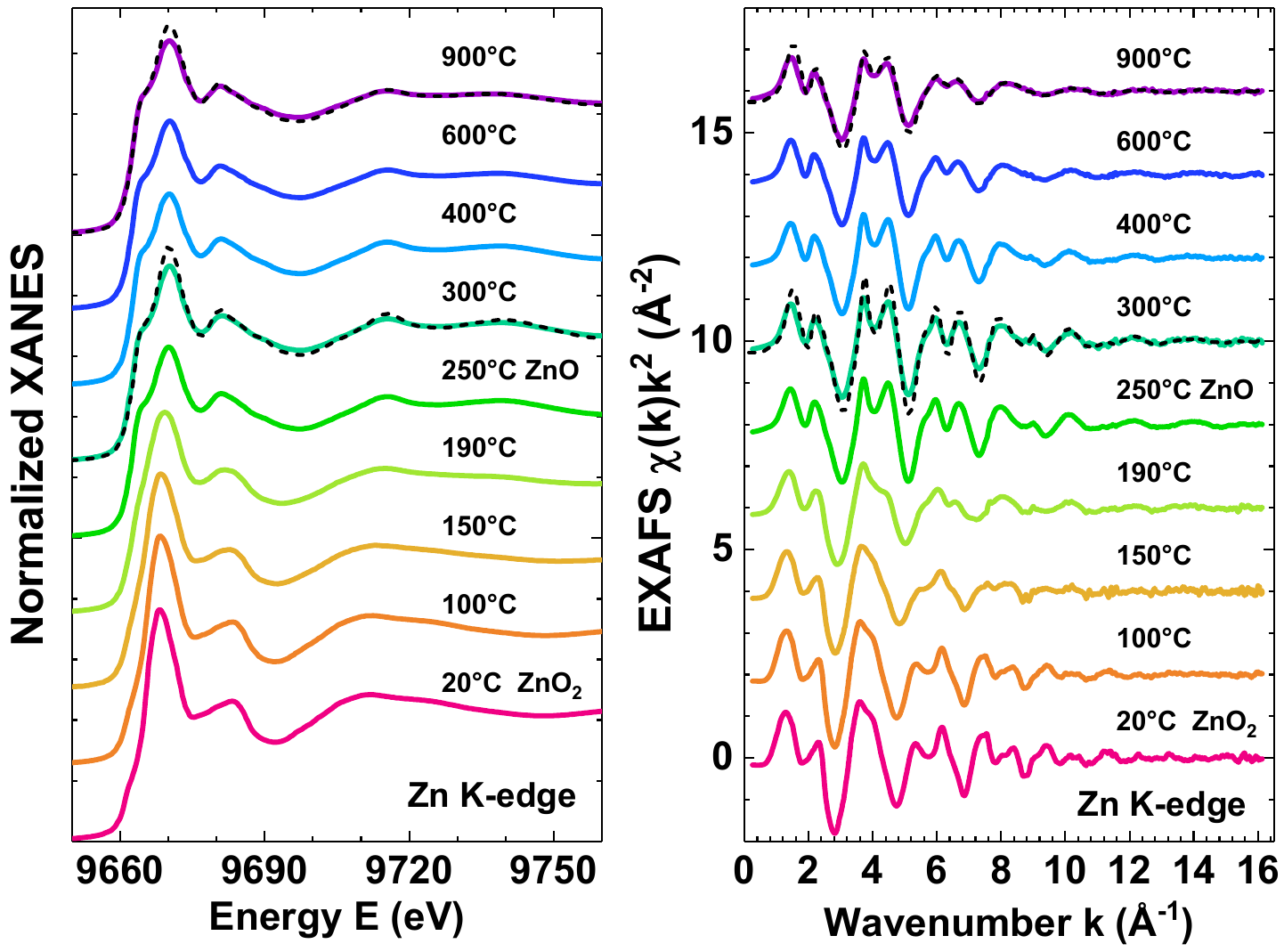}
	\caption{In situ experimental Zn K-edge XANES and EXAFS spectra obtained in the temperature range of 20-900$^\circ$C during the decomposition of ZnO$_2$\ to ZnO. The spectra for reference microcrystalline ZnO are shown by dashed lines at 300$^\circ$C and 900$^\circ$C for comparison.  }
	\label{fig1}
\end{figure*}

\begin{figure*}[h]
	\centering
	\includegraphics[width=0.7\linewidth]{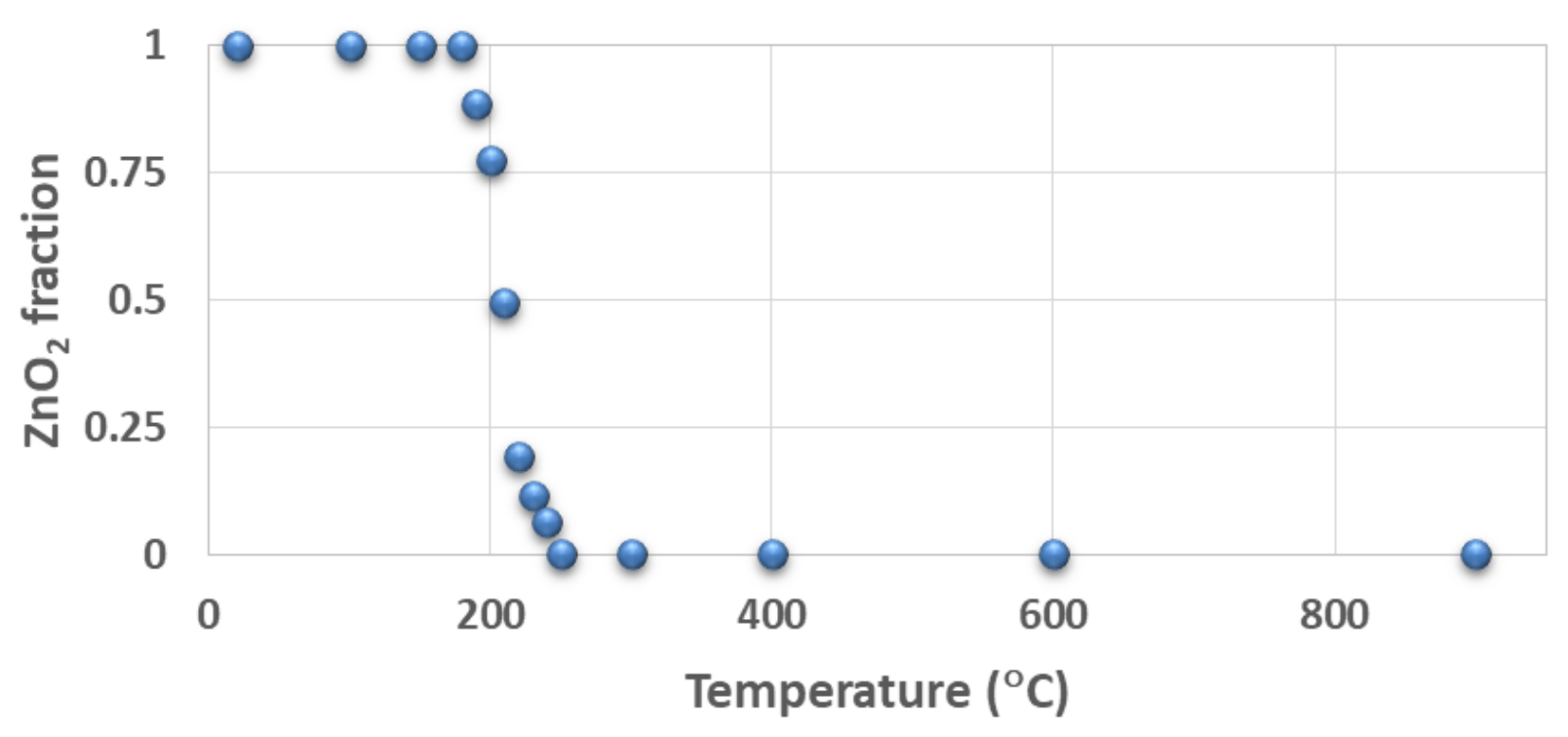}
	\caption{Decomposition process of ZnO$_2$\ to ZnO according to the linear combination analysis of the Zn K-edge EXAFS data.}
	\label{fig2}
\end{figure*}

\medskip
In situ Zn K-edge XANES and EXAFS spectra reported in Figure\ \ref{fig1} indicate the change of the local zinc environment upon heating of ZnO$_2$ from 20 to 900$^\circ$C.  
The principal component analysis of a series of the Zn K-edge EXAFS spectra measured in situ during the ZnO$_2$-to-ZnO decomposition  as a function of temperature suggests that there are only two main components, which are related to the two phases of zinc oxide. Therefore, the linear combination analysis of the EXAFS spectra using the reference data for ZnO$_2$\ and wurtzite-type ZnO allowed us to determine the amount of the two phases upon heating (Figure\ \ref{fig2}). 
Only the peroxide phase exists up to 180$^\circ$C. The decomposition of ZnO$_2$\ occurs in the temperature range 
of 190-240$^\circ$C, where both zinc oxide phases coexist. Finally, only the ZnO phase is present 
above 250$^\circ$C. However, a comparison with the reference spectra of microcrystalline ZnO at 300$^\circ$C and 900$^\circ$C (see dashed lines in Figure\ \ref{fig1}) indicates some expected differences due to the nanocrystalline nature of the decomposition product \cite{Daley2021}.  

\begin{figure*}[h]
	\centering
	\includegraphics[width=0.7\linewidth]{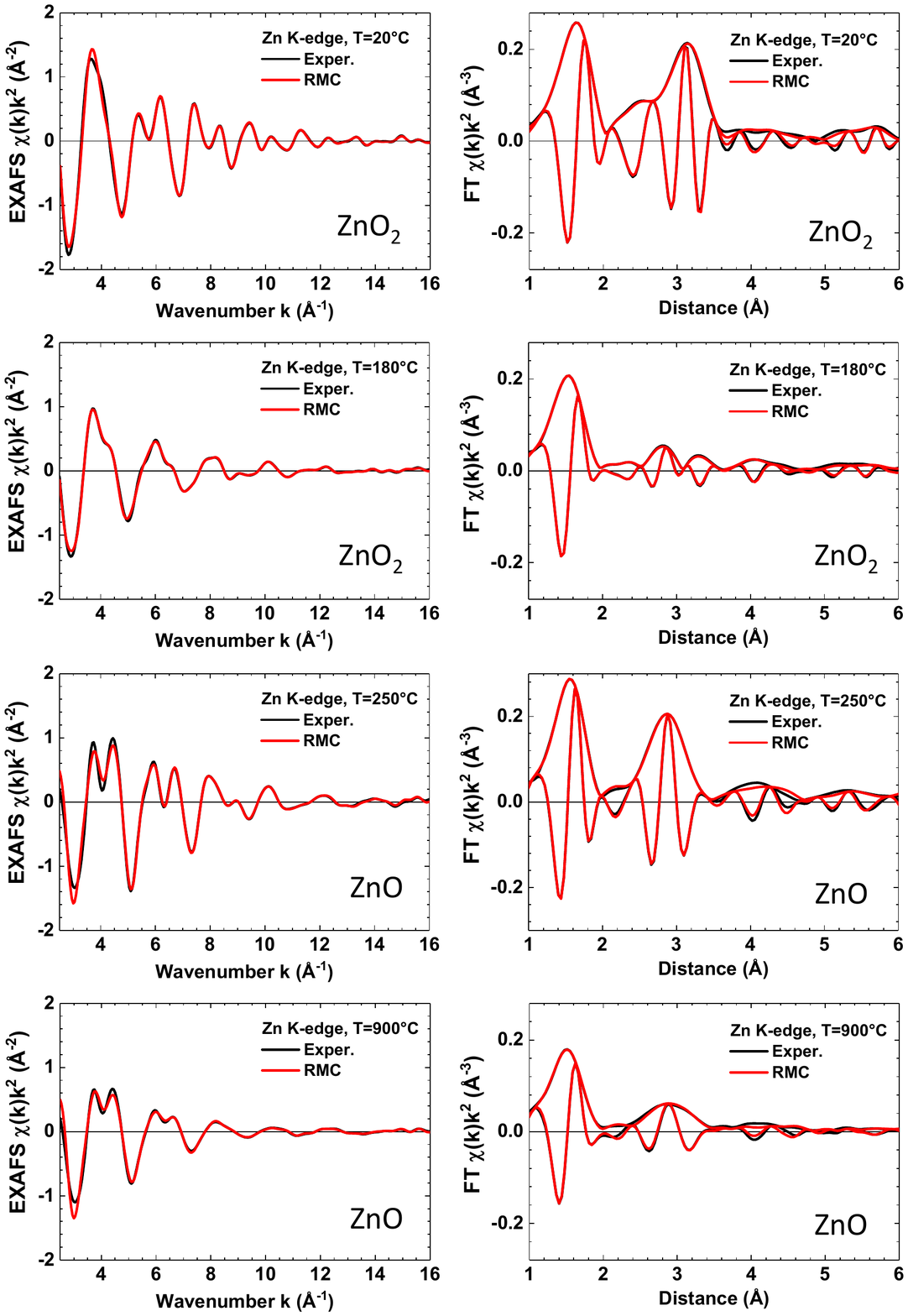}
	\caption{Comparison of the experimental and RMC calculated Zn K-edge EXAFS spectra $\chi(k)k^2$\  
		(left panels) and their Fourier transforms (FTs) (right panels) for selected temperatures. The results for 20 and 180$^\circ$C correspond to the ZnO$_2$\ phase, whereas for 250 and 900$^\circ$C to the ZnO  phase obtained  upon the decomposition of the peroxide.  }
	\label{fig3}
\end{figure*}

\medskip
Based on the results of the PCA/LCA calculations (Figure\ \ref{fig2}), the reverse Monte Carlo simulations of the Zn K-edge EXAFS spectra were performed considering two structural models -- ZnO$_2$\ for temperatures up to 180$^\circ$C and wurtzite-type ZnO for temperatures above 250$^\circ$C.  Both models reproduce well the experimental data at respective temperatures in $k$\  and $R$\ spaces (Figure\ \ref{fig3}). Note that good agreement is observed in the outer coordination shells up to 6~\AA\ with some small deviation for the peak around 4~\AA\ in the Fourier transforms above 250$^\circ$C in the ZnO phase. The observed deviations could be related to the fact that our RMC model with the periodic boundary conditions corresponds to the infinite crystal and neglects structural disorder at the crystallite surfaces that affects stronger the outer coordination shells.  

\begin{figure*}[h]
	\centering
	\includegraphics[width=0.7\linewidth]{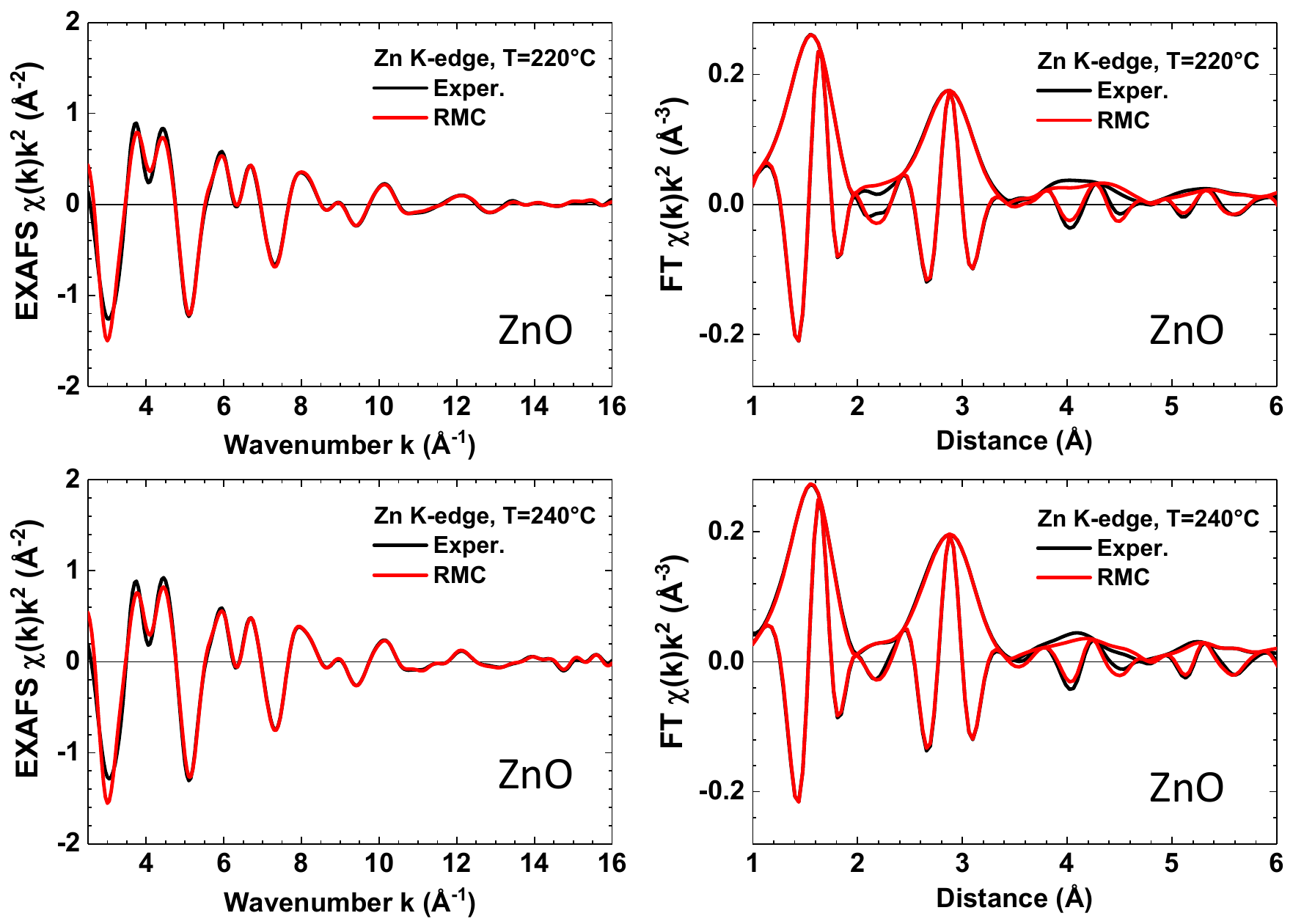}
	\caption{Comparison of the experimental and calculated (RMC) Zn K-edge EXAFS spectra $\chi(k)k^2$\  
		(left panels) and their Fourier transforms (FTs) (right panels) in the intermediate temperature range at 220 and 240$^\circ$C. The RMC simulations were performed using the wurtzite-type ZnO structure model. }
	\label{fig4}
\end{figure*}

\medskip
We have also found that the experimental EXAFS spectra in the intermediate temperature range from 220 to 240$^\circ$C can be well fitted using the wurtzite-type ZnO structure (Figure\ \ref{fig4}) since the expected contribution of the ZnO$_2$\ phase from the PCA/LCA calculations is relatively small. Indeed, the observed agreement is the same as for higher temperatures where only the ZnO phase is present. 

\medskip
Comparison of the EXAFS spectra and their Fourier transforms in Figures\ \ref{fig3} and \ref{fig4} suggests  
their strong dependence on the oxide phase which is related to the difference in the zinc coordination (ZnO$_6$\ vs ZnO$_4$) and the arrangement of the structural units in the crystallographic structures.  While the thermal disorder contributes to the damping of EXAFS oscillations at large $k$-values and the reduction of peaks in the Fourier transforms upon increasing temperature, the variation of the crystallographic structure from ZnO$_2$\ to ZnO produces a dominant effect well observed in the ranges of the first and, especially, second coordination shells (peaks at around 1.5~\AA\ and 2.8~\AA\ in FTs).

\begin{figure*}[h]
	\centering
	\includegraphics[width=0.5\linewidth]{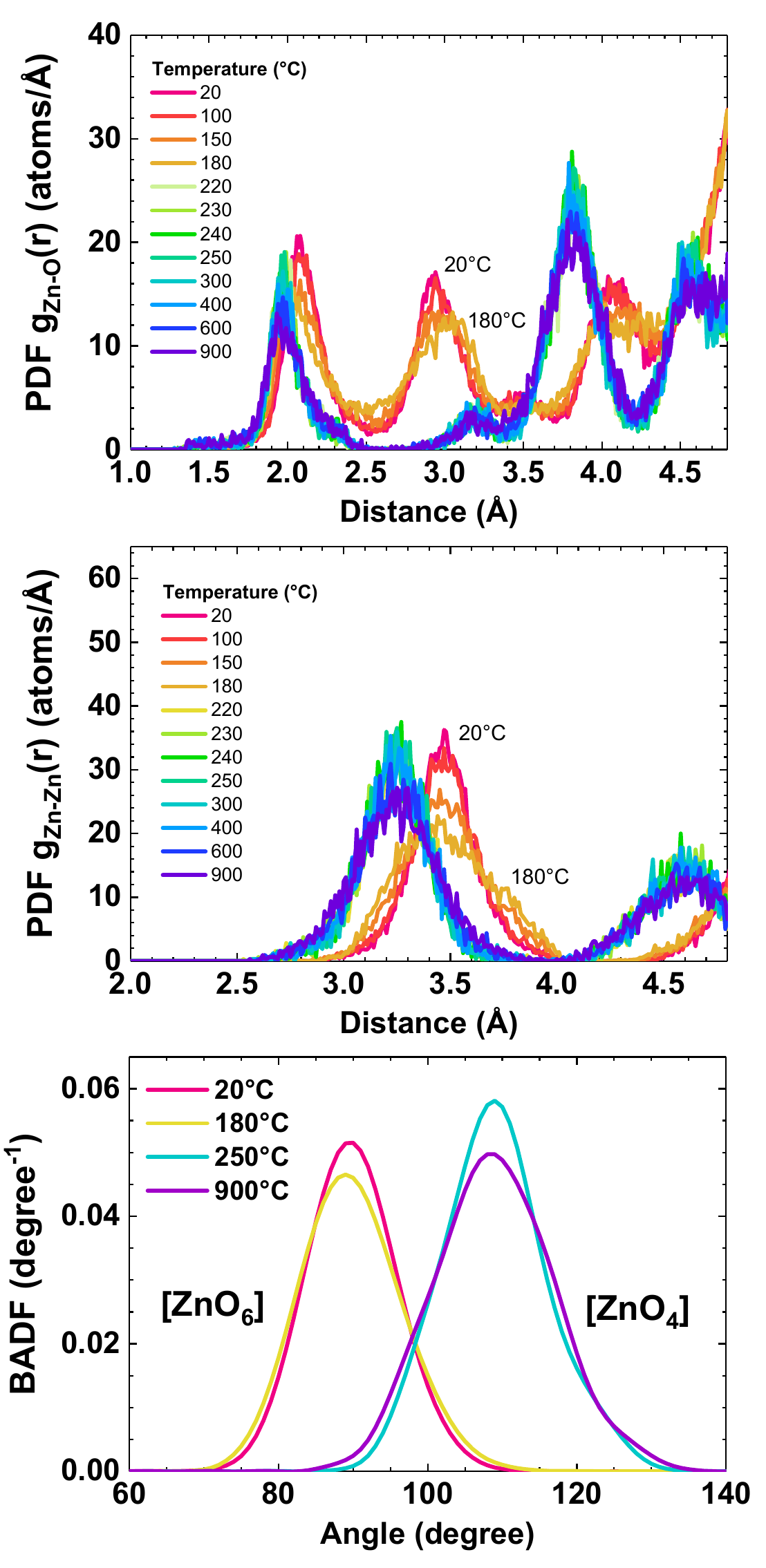}
	\caption{Temperature-dependent atomic pair distribution functions (PDFs) $g(r)$\ for Zn--O and Zn--Zn atom pairs and bond angle distribution functions (BADFs) for the O--Zn--O angles calculated from the results of the RMC simulations of the Zn K-edge EXAFS spectra. }
	\label{fig5}
\end{figure*}

\medskip
Temperature-dependent atomic PDFs  $g(r)$\  for Zn--O and Zn--Zn atom pairs obtained from the results of the RMC simulations are shown in Figure\ \ref{fig5}. They reflect nicely the  evolution of the local atomic environment around zinc atoms. 
Both PDFs in the ZnO$_2$\ phase broaden significantly upon approaching the decomposition temperature when ZnO$_2$\ transforms to nano-ZnO. From 220$^\circ$C, the local environment of the wurtzite-type ZnO phase starts to dominate. However, further  increase in temperature promotes the growth of ZnO crystallites, which starts to compete with the increasing thermal disorder. 
The variation of the local structure in zinc oxides can be also described in terms of the BADFs for the O--Zn--O angles (Figure\ \ref{fig5}). 
The BADFs change as expected from 90$^\circ$\ in ZnO$_6$\ octahedra to about 109$^\circ$\ in ZnO$_4$\ tetrahedra and broaden upon increasing temperature.  

\begin{figure*}[h]
	\centering
	\includegraphics[width=0.7\linewidth]{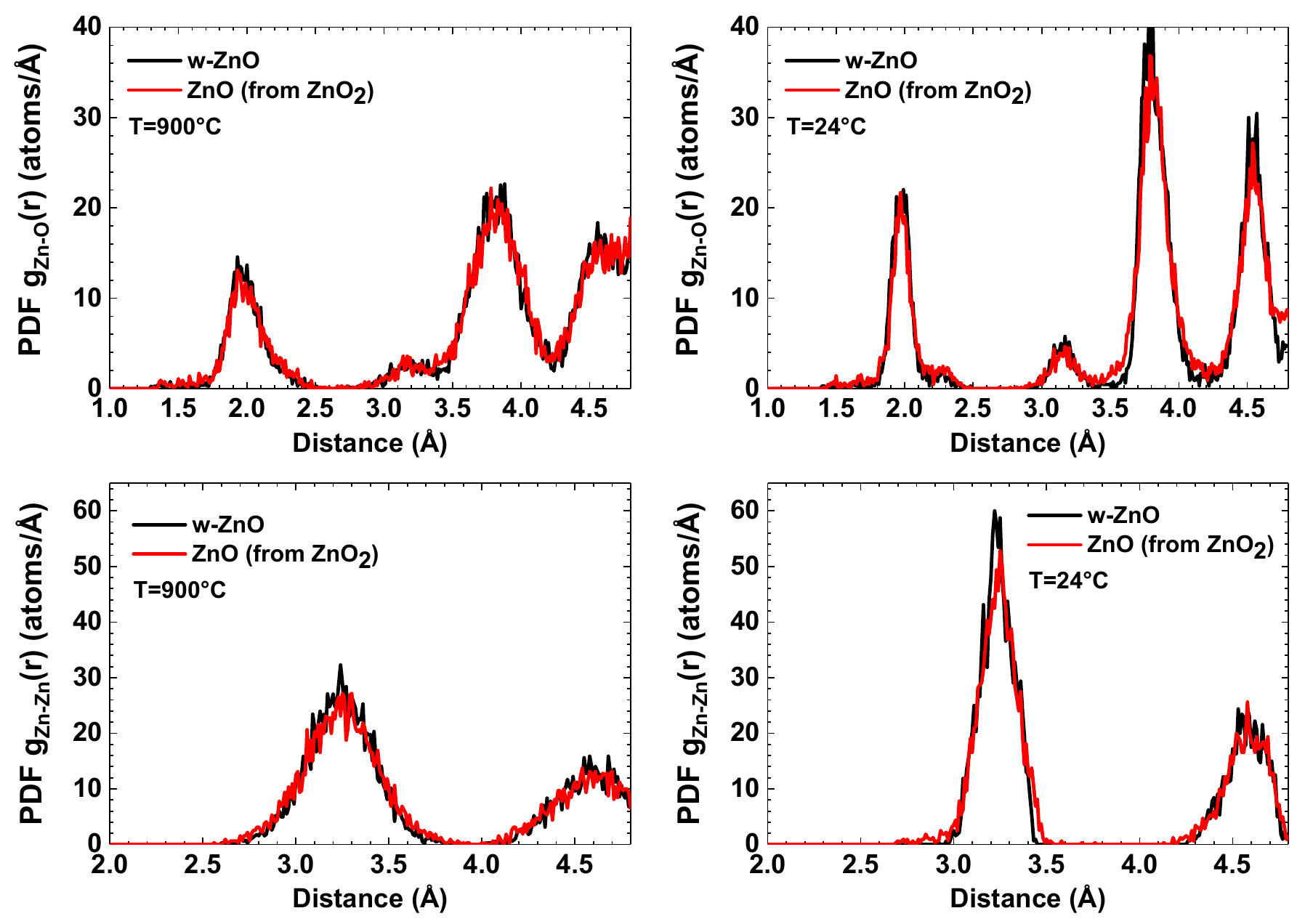}
	\caption{Comparison of the atomic pair distribution functions (PDFs)  $g(r)$\  for Zn--O and Zn--Zn atom pairs at 900$^\circ$C and after cooling down to 24$^\circ$C in w-ZnO and ZnO obtained by decomposition of ZnO$_2$. PDFs were  calculated from the results of the RMC simulations of the Zn K-edge EXAFS spectra. }
	\label{fig6}
\end{figure*}

\medskip
Finally, we compared in \textbf{Figure\ \ref{fig6}} low (24$^\circ$C) and high (900$^\circ$C) temperature PDFs $g(r)$\  for Zn--O and Zn--Zn atom pairs in  microcrystalline wurtzite-type ZnO \cite{Bocharov2021znoht} and the one produced from ZnO$_2$. Their good agreement suggests that the crystallinity of nano-ZnO produced by decomposition almost recovers upon heating to 900$^\circ$C. Nevertheless, some broadening of the outer shell peaks remains in both PDFs after cooling to 24$^\circ$C. 

\begin{figure*}[h]
	\centering
	\includegraphics[width=0.5\linewidth]{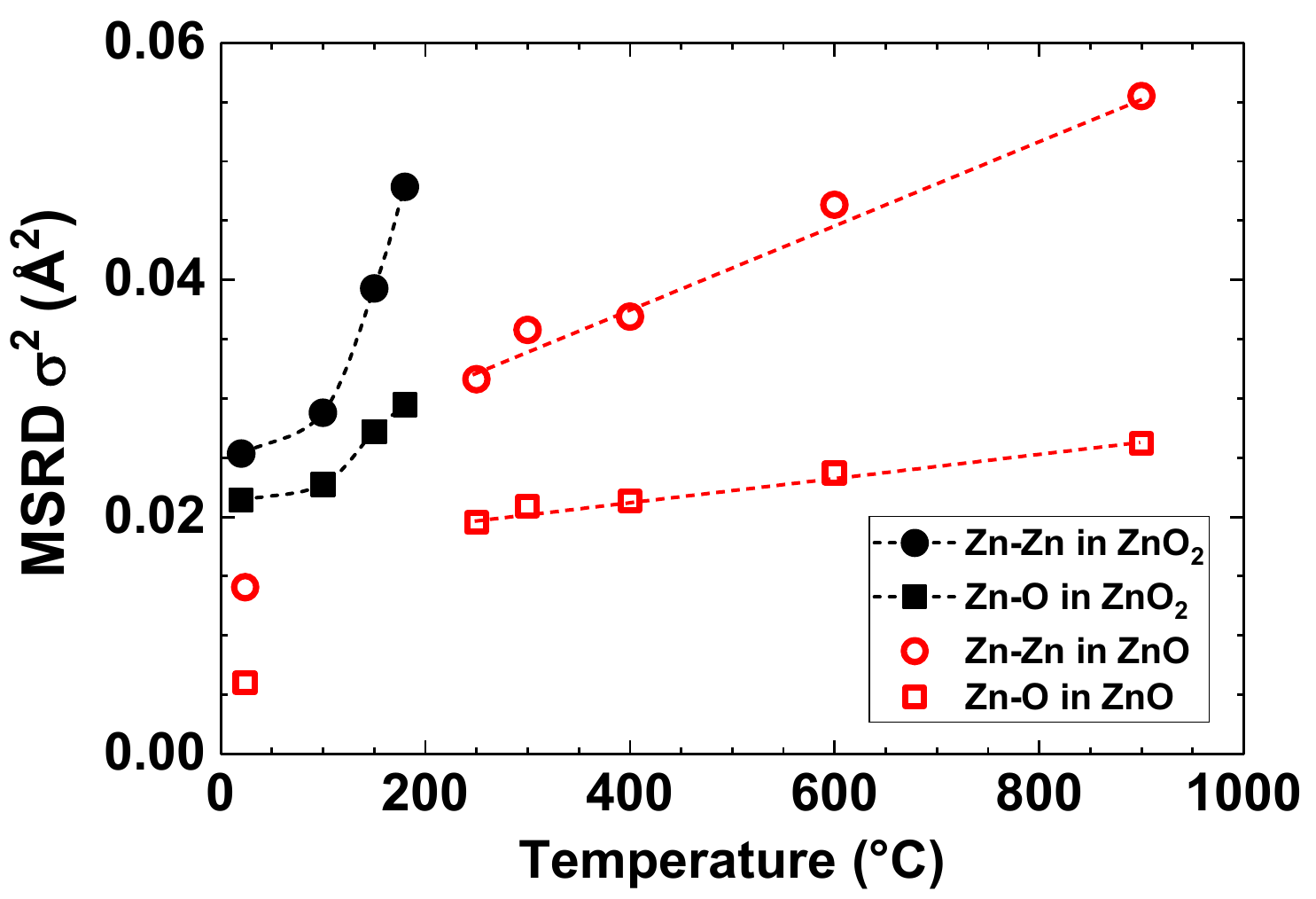}
	\caption{Temperature dependence of the mean-square relative displacements (MSRDs) $\sigma^2$ for Zn--O and Zn--Zn atom pairs in the first and second coordination shells, respectively. 	The dashed lines are guides for the eye. The MSRD values for ZnO at 24$^\circ$C are after cooling.  }
	\label{fig7}
\end{figure*}

\medskip
The PDFs reported in Figure\ \ref{fig5}  were used to extract information on the mean-square relative displacements (MSRDs) for the Zn--O and Zn--Zn atom pairs corresponding to the first and second coordination shells of zinc, respectively (\textbf{Figure\ \ref{fig7}}). The values of MSRDs were evaluated as the second moment of the PDFs in the range of the corresponding peaks.  Strong bonding between Zn and O atoms is responsible for the lower  MSRD $\sigma^2$(Zn--O) values than in the case of the Zn--Zn atom pairs. Both MSRDs $\sigma^2$(Zn--O) and $\sigma^2$(Zn--Zn) increase strongly in ZnO$_2$ before the decomposition temperature, whereas their temperature dependence in the ZnO phase is almost linear. An increase in the size of the ZnO nanocrystals causes a slower than expected growth of the MSRDs with an increase in temperature above 250$^\circ$C. Such behaviour of MSRDs is due to a decrease in static disorder partly compensates  for the increase in thermal disorder. Note that the values of both MSRDs after cooling to 24$^\circ$C\ are close to  that in bulk w-ZnO as supported by the similarity of their PDFs in Figure\ \ref{fig6}.

\section{Conclusions}

\medskip
In situ high-temperature (20-900$^\circ$C) Zn K-edge X-ray absorption spectroscopy study has been performed 
to follow the changes of the local atomic structure during the decomposition of zinc peroxide
to zinc oxide, occurring in the range of 190-240$^\circ$C. 
The use of the principal component and linear combination analyses of the EXAFS spectra allowed us to establish the temperature ranges of the existence and coexistence of the two oxide phases.  
We found that only the peroxide phase exists up to 180$^\circ$C, whereas only the nano-ZnO phase occurs above 
250$^\circ$C.  These conclusions were further supported by the results of the reverse Monte Carlo simulations, which provided us with different distribution functions describing the local environment and its temperature dependence. The calculated atomic pair distribution functions $g(r)$\  for Zn--O and Zn--Zn atom pairs suggest an increase of disorder close to the decomposition temperature. This effect is also well observed in the temperature dependence of the related mean-square relative displacements.
An increase of temperature to 900$^\circ$C promotes the growth of ZnO crystallites. 

\section{Experimental}

\medskip
Nanocrystalline ZnO$_2$\ with the crystallite size of about 22~nm from \cite{Bocharov2022}  and commercial polycrystalline wurtzite-type ZnO (99.99\% purity, Alfa Aesar) were used in the powder form. For high-temperature experiments, each zinc oxide powder was mixed with BN powder and pressed into a pellet.
The amount of zinc oxide was adjusted to give the value of the Zn K-edge (9659~eV) jump $\Delta \mu$\ close to 1.

\medskip
Temperature-dependent Zn K-edge  X-ray absorption experiments were performed in transmission mode on the BALDER beamline located at the 3.0 GeV storage ring of MAX IV Laboratory \cite{Balder2016}. 
X-ray beam from the in-vacuum wiggler source was monochromatized using a liquid-nitrogen-cooled double-crystal Si(111) monochromator and measured by two ionization chambers located before and after the sample and filled with N$_2$\ and Ar gases. The uncoated Si collimating mirror and silica focusing mirror were used for harmonic reduction. The sample was heated in a nitrogen atmosphere using the Linkam furnace  in the temperature range from 20 to 900$^\circ$C and then cooled down to 24$^\circ$C. In situ X-ray absorption spectra were collected at selected temperatures.  Each EXAFS scan took ca. 60 sec, while the temperature was maintained fixed. To assure reproducibility, two repeats were taken for each temperature.

\medskip
\textbf{Acknowledgements} \par 
The authors wish to thank  Dr.phys. R. Kalendarev for the synthesis of ZnO$_2$\ sample. 
A.K. would like to thank the financial support of the ERDF project No. 1.1.1.1/20/A/060.
The experiment at the MAX IV synchrotron was performed within the project 20190823.
Institute of Solid State Physics, University of Latvia as the Center of Excellence has received funding from the European Union's Horizon 2020 Framework Programme H2020-WIDESPREAD-01-2016-2017-TeamingPhase2 under grant agreement No. 739508, project CAMART2.

\medskip
\textbf{Conflict of Interest}
The authors declare no conflict of interest.

\medskip
\textbf{Data Availability Statement}
The data that support the findings of this study are available from the
corresponding author upon reasonable request.

\medskip

\bibliographystyle{MSP}
\bibliography{zno2.bib}


\newpage

\begin{figure}
\textbf{Table of Contents}\\
\medskip
  \includegraphics[width=0.5\linewidth]{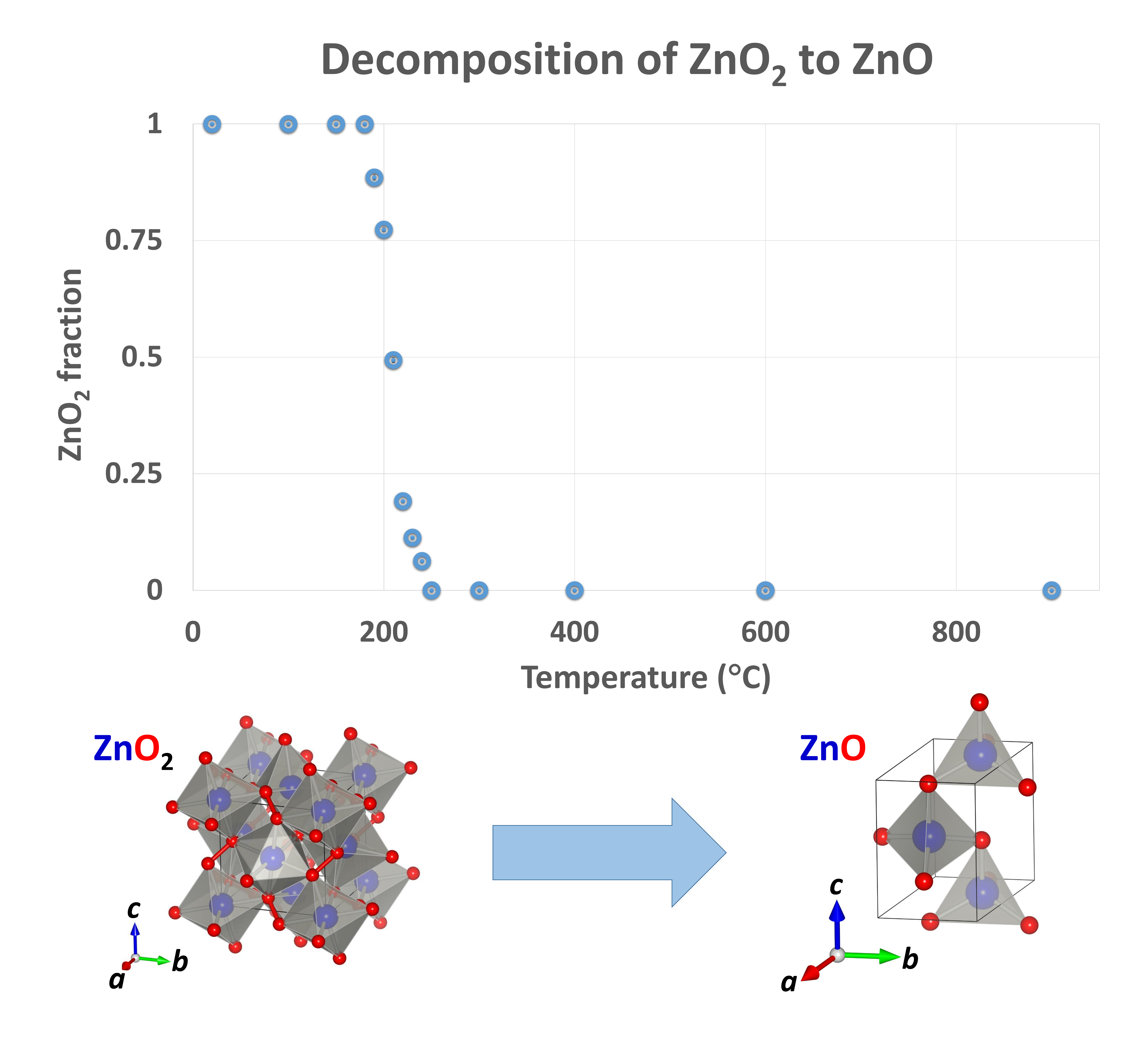}
  \medskip
  \caption*{In situ Zn K-edge X-ray absorption spectroscopy combined with the reverse Monte Carlo
  	simulations was used to study the decomposition process of ZnO$_2$\ to ZnO. 
  	The variations in the local environment around zinc atoms are discussed based on the analysis of pair atomic and bond angle distribution functions. }
\end{figure}

\end{document}